\documentclass[11pt,twoside,a4paper]{article}

\usepackage[american]{babel}
\usepackage{bm}
\usepackage{amsfonts}
\usepackage{graphicx}
\usepackage{amsmath}
\usepackage{fancyhdr}
\usepackage{a4wide}

\usepackage{setspace}
\usepackage[utf8]{inputenc}

\usepackage{amsthm} 

\usepackage{mathtools}

\DeclarePairedDelimiterX\braket[2]{\langle}{\rangle}{#1 \delimsize\vert #2}

\usepackage{siunitx} 
 \usepackage{movie15}
\usepackage{hyperref}
\usepackage{booktabs}

\usepackage{cite}

\usepackage{enumitem}
\usepackage{amssymb}

\newcommand{\disp}{\displaystyle}

\usepackage{color}

\newcommand{\zg}{\gamma}
\newcommand{\zl}{\lambda}
\newcommand{\zm}{\mu}

\newcommand{\zo}{\omega}
\newcommand{\zr}{\rho}

\newcommand{\zy}{\psi}

\newcommand{\zh}{\eta}

\newcommand{\dif}{\; \textrm d }

\newcommand{\dsp}[2]{\frac{  \partial^2 #1 }{  \partial #2^2   } }
\newcommand{\dst}[2]{\frac{  \dif^2 #1 }{  \dif #2^2   } }
\newcommand{\dpp}[2]{\frac{  \partial #1 }{  \partial #2   } }
\newcommand{\dpt}[2]{\frac{  \dif #1 }{  \dif #2   } }

\theoremstyle{definition}

\renewcommand{\endproof }{\hfill $\square$\\ }

\newtheorem{theorem}{Theorem}

\newtheorem{lemma}{Lemma}

\begin{document}

\begin{center}
\Large{\textbf{A Wigner–Lindblad phase space model for open quantum dynamics }}\\
	\small{O. Morandi}	\\
\vskip0.5cm
	\textit{\textsf University of Florence  \\
		Department of Mathematics and  Computer Science U. Dini, \\
Firenze, Italy}	
\vskip0.5cm
	\textit{omar.morandi@unifi.it}
\end{center}
\begin{abstract}
	We investigate the collisional quantum dynamics of an open system in contact with a thermal bath. We propose a particle-bath interaction model based on the Weyl quantization principle and the Wigner phase space description of quantum evolution. By factorizing the reduced Hamiltonian of the open system via supersymmetric quantum mechanics techniques, we construct system-dependent creation and annihilation operators for arbitrary potential landscapes. This allows us to extend the classical Fokker--Planck relaxation mechanism to incorporate Lindblad quantum stochastic jumps within the Wigner phase-space formalism. Through a semiclassical expansion of the Moyal star product to second order, we obtain a generalized Wigner--Lindblad kinetic model featuring position-dependent friction and diffusion terms. We establish the mathematical well-posedness of the Lindblad problem and we validate our approach through numerical simulations. Our model resolves unphysical long-time artifacts of standard quantum Fokker--Planck models in quartic double-well traps, accurately captures dissipative quantum state localization in 2D molecular potentials (4D phase space), and effectively evaluates electron transport and trapping times in InAs/InP resonant tunneling diodes.
\end{abstract}


\section{Introduction }
 Understanding the interactions between quantum systems and their environments is crucial for developing stable quantum technologies and accurate physical models of open quantum dynamics. The environment, such as a surrounding radiation field or an atomic and molecular reservoir, is typically modeled as an infinite collection of harmonic oscillators that induce decoherence, dephasing, and time-irreversible relaxation toward thermal equilibrium.
 Simulating open quantum systems by directly solving Schrödinger's equation for the combined system and environment is usually infeasible due to the exponential growth of the Hilbert space, which scales dramatically with the number of environmental degrees of freedom.
 Conventionally, system-environment interactions are treated in the weak-coupling regime, where bath memory times are assumed to be negligible compared to the characteristic system timescales. In this regime denoted as  Markovian limit, the evolution of the reduced density matrix is prominently formulated via the Gorini--Kossakowski--Sudarshan--Lindblad (GKLS) master equation framework.
 Driven by the need to capture strong-coupling and finite-temperature effects, the original GKLS framework has been extended in several directions. Modern approaches to open systems, such as tensor network techniques, the hierarchical equations of motion, and the quantum hierarchical Fokker-Planck equation, have been developed to capture complex dynamical regimes and enable large-scale or stochastic simulations \cite{Grossmann_14,Luchnikov_19,Tanimura_20,Takahashi_24,Sander_25}. In parallel, substantial effort has been devoted to modeling and simulating non-Markovian environments \cite{Breuer_16,Cygorek_22,Gribben_22}.
 As an alternative to direct density matrix propagation, the steady states and temporal evolution of open systems can be efficiently obtained via stochastic unravelings, such as quantum-jump trajectory wave-function Monte Carlo methods \cite{Johansson_12,Sander_25}, as well as imaginary-time evolution algorithms adapted to Lindbladian dynamics \cite{Kamakari_22}. Contemporary approaches also leverage artificial neural networks to parameterize open quantum states and capture complex environmental interactions \cite{Vicentini_19,Reh_21}.

The quantum Wigner formalism of phase space quasi-probability distribution functions represents an equivalent alternative to the density matrix formalism for describing the dynamics of mixed quantum systems. One of the defining characteristics of the Wigner framework is its close resemblance to the statistical description of particle dynamics in classical physics.
Introductory overviews of the Wigner formalism, its mathematical foundations, and its applications can be found in several reviews \cite{Lee_95,Schleich_01,Case2008,Polkovnikov_10,De_Gosson_17,Revzen_06}. The Wigner framework encompasses a broad spectrum of applications, accommodating both continuous and discrete degrees of freedom \cite{Agarwal_81,Varilly_89,Brif_98,Tilma_16,Klimov_17}.
Since its introduction, the Wigner formalism has become a well-established framework across various research fields, including solid-state physics \cite{Bordone_99,Demeio_02,Unlu_04,Sverdlov_05,Weinbub_18,Bordone_19,Camiola_21,Fillion2022}, graphene and nanomaterials \cite{morandi_JPA_11,Figueiredo_22,Tarasov2021,Tahir2013,Jauho2011}, quantum optics \cite{Dragoman_05,Dittrich_06,Alonso_11,Mout_18,Wuv_22,Morandi_JPA_24,Morandi_KRM_25}, optically controlled cold atoms \cite{Saffman_19,Bluvstein_23,Zhang_15,Manfredi_17,Hwang_23,Cicali_24,Morandi_25_PRA,Browaeys_20}, Bose-Einstein condensates \cite{Griffin_book,Choi_12,Mor_14PRA}, and molecular dynamics \cite{Wen_22,He_24}.
The Wigner formalism has also proven to be an effective tool for investigating fundamental theoretical aspects of physics, such as the classical limit of quantum dynamics \cite{Markowich_89,Markowich_Ringhofer,Lions_93,Sparber_03,Gat_14,Morandi_26_arXiv,Morandi_KRM_25} and local quantum equilibrium \cite{Gardner_96,Degond_05,Romano_07,Hurst_14}.
More recently, the Wigner formalism has been applied to quantum optimal control \cite{Chen_11,Rosi_13,Yu_26,Morandi_24_SIAP} and to the study of particle--particle correlations and quantum entanglement \cite{Morandi_24_PLA,Morandi_26_PRA,Zawu_26}.
Over the last few decades, several numerical schemes for solving the Wigner equation have been developed. Deterministic algorithms rely on the direct discretization of the Wigner equation \cite{Ringhofer1990,ArnoldRinghofer1996,Frensley,ShaoLuCai2011,Cabrera_15,Chai_Morandi2015,DordaSchuerrer2015}, whereas stochastic algorithms based on Monte Carlo methods reconstruct phase space dynamics via sampling techniques \cite{JacoboniBertoniBordoneBrunetti2001,JacoboniBordone2004,Muscato_16,Weinbub_18}.

The connection between the quantum Wigner-Fokker-Planck equation and the Lindblad formalism is well known and has be investigated both form a theoretical \cite{Arnold_08} and a modeling view point \cite{Oliveira_23}  and the potential application to continuous measurement and control protocols \cite{Annby-Andersson_22}.  
In this paper, we propose a novel heuristic model to describe relaxation processes in open quantum systems. Our approach extends the quantum Wigner-Fokker-Planck phase space framework by assuming that the particle-bath interaction is governed by the Lindblad master equation. We show that the standard quantum Wigner-Fokker-Planck equation is recovered as a special case when the Weyl quantization procedure is expanded to second order in the $\hbar$-semiclassical expansion of the dynamical equations and the potential is assumed to be harmonic. Selecting appropriate environment interaction operators is a crucial aspect of our method. We apply a Hamiltonian factorization technique that uniquely defines a pair of creation and annihilation operators associated with each specific potential profile.

The paper is organized as follows: In Section~\ref{Sec_model}, we introduce the general framework of open quantum systems and motivate the need for a phase-space relaxation model beyond standard harmonic approximations. In Section~\ref{sec_well_pos}, we establish the mathematical well-posedness of the Lindblad dynamics. In Section~\ref{sec_WL_eq}, we perform a second-order semiclassical expansion full quantum equations and we derive the generalized Wigner--Lindblad equation. In Section~\ref{sec_num_res}, we test the physical validity our model across three test cases: an anharmonic double-well trap, a two-dimensional molecular potential in a four-dimensional phase space, and an InAs/InP double-barrier resonant tunneling diode.

 \section{Model}\label{Sec_model}

In open quantum systems theory, the time evolution of the density matrix is required to be governed by a completely positive, trace-preserving Markovian semigroup, often referred to as a Markovian quantum dynamical semigroup. We denote by $\zr$ the mixed state described by the reduced density matrix of a particle system interacting with an external environment, referred to as a thermal bath. The Gorini--Kossakowski--Sudarshan--Lindblad (GKSL) framework ensures that the most general dynamical generator describing the evolution of an open quantum system can be expressed as \cite{Davies_69,Gorini_76,Lindblad_77,Davies_77,Frigerio_82,Alicki_87}
\begin{align}\hspace*{-20pt}
	\text{(LP)}: \;	\left\{ 	
	\begin{array}{l}
		\disp\hbar \dpp{\zr}{t} = - i  \left[H,\zr\right] + \zg^- \left( a\zr a^\dag -\frac{1}{2} \left\{ a^\dag a , \zr \right\}  \right)  + \zg^+ \left( a^\dag \zr a -\frac{1}{2} \left\{ a a^\dag , \zr \right\}  \right)  \\[4mm]
		\left. \zr \right|_{t=0} = \zr_0
	\end{array}\label{Lind_dens_mat02} 
	\right.
\end{align}
where $\zg^+,\zg^->0$ with $\zg^+< \zg^-$, and the initial state $\zr_0$ belongs to $ \mathcal{J}_1^+ (L^2(\mathbb{R}))$, the cone of positive trace-class operators on $L^2(\mathbb{R})$. In what follows, we refer to Eq.~\eqref{Lind_dens_mat02} as the Lindblad Problem (LP).

For simplicity, we restrict our analysis to a single pair of operators, $a$ and $a^\dagger$, modeling the system--environment interaction. The extension to a countable set of operators is straightforward. Physically, the Lindblad master equation can be interpreted as describing the ensemble-averaged dynamics of a stochastic process, wherein the operators $a$ and $a^\dagger$ act randomly on the quantum system with characteristic rates $\zg^+$ and $\zg^-$, respectively. The term $a \rho a^\dagger$ represents discrete Markovian quantum jumps, whereas the anticommutator terms account for the continuous, non-Hermitian evolution during no-jump intervals, thereby ensuring trace preservation and proper state normalization. A vast body of literature addresses this subject, covering both theoretical modeling and rigorous mathematical studies (see, e.g., \cite{Gribben_22,Plenio_98,Frohlich_25}).

The most common choice for describing the interaction relies on identifying the operator $a$ and its formal adjoint $a^\dag$ with standard photon annihilation and creation operators. However, extending the validity of these operators beyond quantum optical systems may lead to unphysical results or violate fundamental principles, as will be discussed in detail below.
We formulate heuristic criteria to guide the selection of appropriate Lindblad operators for a given physical context. Due to the complexity of the problem, we focus on constructing well-defined approximate Lindblad operators that satisfy essential requirements regarding energy dissipation and thermalization toward equilibrium. Our approach, which generalizes the algebraic construction of ladder operators originally developed for the harmonic oscillator, provides a computationally tractable framework for modeling relaxation, dispersion, and decoherence phenomena in realistic open quantum systems.

To illustrate the method, we consider the $n$-dimensional harmonic Hamiltonian $
H_{\text{Har}}= - \frac{\hbar^2}{2m} \Delta_x  +\sum_{i=1}^n\frac{\zo_i^2 }{2} x_i^2 	$, where $\zo\in \mathbb{R}^n$ is the vector of oscillator strengths. In this context, the standard annihilation and creation operators are defined as $
a =    \frac{\hbar}{\sqrt{2m}}\nabla_x  + \frac{1}{\sqrt{2}}\zo \circ x $, and $
-  \frac{\hbar}{\sqrt{2m}}\nabla_{x}  + \frac{1}{\sqrt{2}}\zo \circ x$,
respectively,  where $\zo \circ x \in \mathbb{R}^n$ denotes the vector with components $(\zo \circ x)_i = \omega_i x_i$. The Hamiltonian can be expressed in terms of these operators as $
H_{\text{Har}} = a^\dag	a +E_0 $, with ground-state energy $E_0 = \frac{\hbar}{2\sqrt{m}} \sum_{i=1}^n \zo_i $. 
The underlying microscopic model representing the system--bath interaction assumes that the environment randomly interacts with (or weakly measures) the system, applying the jump operators $a$ and $a^\dagger$ to the reduced density matrix operator. We emphasize that $a$ and $a^\dagger$ act exclusively on the Hilbert space of the reduced system rather than on the bath degrees of freedom. For a particle bound within a harmonic trap, the annihilation operator $a$ corresponds to the extraction of exactly one quantum of energy $\hbar \omega$ from the system to the environment, inducing the state transition $\psi_i \rightarrow \psi_{i-1}$ (for $i > 0$), where the $\psi_i$ represent the eigenstates of the harmonic oscillator. Beyond the harmonic potential, however, this simple correspondence between ladder operators, discrete energy quantum jumps, and the direction of energy flux no longer holds.

Inspired by a well-established operator decomposition technique developed within Supersymmetric Quantum Mechanics (SUSY QM), we determine a pair consisting of a non-Hermitian operator and its adjoint, $(a,a^\dag)$, for quantum-jump interactions such that the Hamiltonian can be factorized as $ H=a^\dag a+I E_0$ \cite{Sukumar_85,Cooper_95}. More precisely, given a system subject to a general external potential $U(x)$ with Hamiltonian $H=-\frac{\hbar^2}{2m}\Delta_x+U$, we require that the annihilation operator $a$ satisfies
\begin{align}
	&\text{Hamiltonian factorization }\quad  H=a^\dag a+I E_0\label{Ham_fact}\\ 
	&\text{Annihilation of the ground state }     \quad a|\psi_0\rangle = 0\label{annih_gr_st}
\end{align}
where $\zy_0$ denotes the ground state and $E_0$ the ground-state energy.
According to SUSY theory, the preceding equations can be solved using the following ansatz
\begin{align}
	a &=\frac{\hbar}{\sqrt{2m}} \nabla_{x}  + W(x) \; , \label{ansatz_a}
\end{align}
where $W :\mathbb{R}^n\rightarrow \mathbb{R}^n$ is a real field. Eq. \eqref{ansatz_a} yields
\begin{align*}
	a^\dag a = &-\frac{\hbar^2}{2m}\Delta_x  + |W|^2 - \frac{\hbar}{\sqrt{2m}} \text{div}_x W =H-U  + |W|^2 - \frac{\hbar}{\sqrt{2m}} \text{div}_x W \;.
\end{align*}
Equation \eqref{Ham_fact} is satisfied if the following condition holds
\begin{align}
	&  |W|^2 - \frac{\hbar}{\sqrt{2m}} \text{div}_x W =   U(x)-E_0\;. \label{W_riccati}
\end{align}
Correspondingly, Eq. \eqref{annih_gr_st} yields $W(x) \zy_0 = -\frac{\hbar}{\sqrt{2m}} \nabla_{x} \zy_0 $, or equivalently $W(x) =  -\frac{\hbar}{\sqrt{2m}} \dpt{\ln \big(\psi_0(x)\big)}{x} $. It is straightforward to verify that Eq. \eqref{annih_gr_st}, together with the ansatz in Eq. \eqref{ansatz_a}, implies Eq. \eqref{W_riccati}. With this choice, the fundamental commutation relations take the form
\begin{align}
	[a^\dag, a] = & - \hbar \sqrt{\frac{2}{m}} \dpt{W}{x} \label{comm_adag_a}\\
	\left[ a,  H \right]  =& \hbar \sqrt{\frac{2}{m}}  \; \dpt{W}{x} \;   a \;.
\end{align}
For the harmonic Hamiltonian, the ground state is $\zy_0 = C e^{-\frac{\sqrt{m}}{2\hbar }\sum_{i=1}^n \zo_i x_i^2 }$, where $C$ is a normalization constant, corresponding to $W= \frac{(\zo \circ x)  }{\sqrt{2}} $, thereby recovering the standard harmonic annihilation and creation operators. Although the preceding relations (Eq. \eqref{annih_gr_st} and the ansatz in Eq. \eqref{ansatz_a}) are well established within the SUSY framework, they have not yet been utilized to construct collisional operators in the context of open quantum systems. 

It is useful to consider the energy flux associated with the LP. By straightforward calculations, the Lindblad master equation yields the following expression for the time evolution of the expected energy of the state $ \langle H \rangle = \text{Tr}(\rho H)$ 
\begin{align*}
	\dpt{\langle H \rangle}{t} =&     \sqrt{\frac{2}{m}}    \left[-  \zg^- \text{Tr}\left\{ \rho \,  a^\dag \; \dpt{W}{x} \;   a  \right\}+ \zg^+ \; \Re \; \text{Tr}\left\{ \rho \,     a a^\dag   \dpt{W}{x}  \right\}\right]\\
	=&       \sqrt{\frac{2}{m}}  \sum_\zl p_\zl  \bigg( -(\zg^- - \zg ^+) \int   \dpt{W}{x} \left|a  \zy_\zl (x) \right|^2  \dif x \\
	& + \zg^+    \chi \int \left(2 \left|\dpt{W}{x}   \right|^2+ W \dst{W}{x}   - \frac{\chi}{2}   \frac{\dif^3 W}{\dif x^3} \right) |\zy_\zl |^2  \dif x   \bigg) \;.
\end{align*}
To analyze the physical consistency of the energy flux, we consider the case where the energy gain term is zero, $\zg^+=0$. In this scenario, we expect the energy flux to be directed strictly from the system to the bath. Consequently, we should observe a monotonic energy decrease for all states except the ground state. From the previous equation, we obtain 
$
\dpt{\langle H \rangle}{t} =  -  \zg^-  \sqrt{\frac{2}{m}}  \sum_\zl p_\zl  \int   \dpt{W}{x} \left|a  \zy_\zl (x) \right|^2  \dif x 
$.
This shows that the energy loss term is only locally well-defined, i.e., provided that $ \dpt{W}{x} \geq 0$. This condition plays a crucial role in the proof of well-posedness of the LP.

\section{Well posedness of the Lindblad problem}\label{sec_well_pos}

In this section, we focus on the mathematical well-posedness of the Lindblad problem defined in Eq.~\eqref{Lind_dens_mat02}. A primary motivation for a rigorous mathematical study of the Lindblad equation stems from the fact that, in general, Lindblad semigroups associated with unbounded operators, such as creation and annihilation operators, are not necessarily conservative, which can lead to a loss of probability mass in finite time. 
The exact characterization of Lindblad dynamics driven by unbounded generators remains an open problem. 
Sufficient conditions ensuring the well-posedness of Lindblad dynamics have been systematically developed across a series of works establishing criteria that guarantee the Markovianity of the process~\cite{Chebotarev_Fagnola_98}, the existence of stationary states~\cite{Fagnola_Rebolledo_01}, and the ergodicity or long-time behavior of the system~\cite{Fagnola_Rebolledo_98b, Fagnola_08}. Introductory overviews of these mathematical aspects and the Lindblad problem are provided in~\cite{Fagnola_99, Daniel_20}. Relevant applications of these theoretical results can be found in~\cite{Bahn_06}, while further developments are discussed in~\cite{Androulakis_15, Siemon_17}.

For simplicity, in this section we restrict our analysis to the one-dimensional (1D) case. The following theorem establishes the well-posedness of the LP.
\begin{theorem}\label{teor_well_pos-LP}
	Under the assumptions: $i)$ $U = U_1 + U_2$ with $U_1 \in L^2(\mathbb{R})$, $U_2 \in L^\infty(\mathbb{R})$
	and $ii)$ $W \in L^\infty(\mathbb{R})\cap H^2_{\text{loc} }(\mathbb{R})$, there exist $M ,M_1> 0$ such that $ 0\leq \dpt{W}{x}  < M  $, $\left|\dst{W}{x}\right|  < M_1 $, 
	the Lindblad problem \eqref{Lind_dens_mat02} generates a conservative $w^*$-continuous semigroup $\mathcal{T}_{t\ge0} : \mathcal{J}_1^+(L^2(\mathbb{R})) \to \mathcal{J}_1^+(L^2(\mathbb{R}))$ with  a unique invariant state. 
\end{theorem}
We recall that an evolution semigroup is said to be conservative if it preserves the trace of the density matrix, i.e., $\text{Tr}(\rho) = 1$ implies $\text{Tr}(\mathcal{T}_t(\rho)) = 1$ for all $t \ge 0$. For the formal definition of weak-star ($w^*$) convergence, we refer the reader to \cite{Fagnola_Rebolledo_03} and Appendix~\ref{App_proof_Th_wel_pos}.

At first glance, the assumption concerning the sign of the derivative of $W$ may appear a little artificial or overly restrictive. However, a comparison with the Fokker--Planck model, obtained as a special case for harmonic confinement (see Sec.~\ref{sec_WL_eq}), reveals the difficulties and critical issues associated with a negative derivative of $W$. A negative derivative of $W$ corresponds to negative values of the friction coefficient in the Fokker--Planck operator, which in general can induce instability and prevent the existence of stationary states due to the onset of runaway solutions or limit cycles. In general, the existence of a stationary solution requires the absence of limit cycles and guarantees that the positive friction terms dominate the dynamics, leading to asymptotically stable solutions.
 



\proof \emph{of Th \ref{teor_well_pos-LP}}: 
To establish notation, we define the operator $
a = \frac{\hbar}{\sqrt{2m}} \frac{\mathrm{d}}{\mathrm{d}x} + W(x) $ 
with domain $D(a) = H^1\left(\mathbb{R}\right)$, and the Hamiltonian $H = -\frac{\hbar^2}{2m}\frac{\mathrm{d}^2}{\mathrm{d}x^2} + U(x)
$, with domain $D(H) = H^2\left(\mathbb{R}\right)$, where $H^1$ and $H^2$ denote the standard Sobolev spaces. 
Under assumption $i)$, $H$ is self-adjoint. The proof of the existence of the conservative $w^*$-continuous semigroup relies on a direct application of the results in~\cite{Chebotarev_Fagnola_98}. Details of the proof of the first part of the theorem are given in Appendix~\ref{App_proof_Th_wel_pos}. 
Here, we focus on the existence of invariant states by applying the results of~\cite{Fagnola_Rebolledo_01}. The main difficulty lies in establishing a Lyapunov-like inequality based on self-adjoint operators $X \ge 0$ and $Y$ bounded from below, such that 
$
A \doteq 2\text{Re}\langle (G_0+H)u, X u \rangle +\zg^- \langle a u, X a u \rangle +\zg^+ \langle a^\dag u, X a^\dag u \rangle \leq - \langle u, Y u \rangle 
$ 
for all bounded operators $u \in B(L^2(\mathbb{R}))$. The operator $G_0$ is defined in Eq.~\eqref{def_G} of Appendix~\ref{App_proof_Th_wel_pos}. We set
$
X= - 2 G_0 =  \zg^- a^\dagger a + \zg^+ a a^\dagger $ and $
Y = -   (2\chi \zg^-)^2 \sup_{x\in\mathbb{R}} \left(W^2+ \left| \dpt{W}{x}\right|^2+\left| \dst{W}{x}\right|^2 \right)   I $.
Using the commutator relation~\eqref{comm_adag_a}, a straightforward calculation yields
\begin{align*}
	A 
	=	&  2\chi\left( \zg^-  +  \zg^+ \right)   \int\left(-\zg^- | a u|^2 +\zg^+ | a^\dag u|^2 \right) \dpt{W}{x}   \dif x   - 4\chi^2   \zg^+\zg^-    \int W   \dst{W}{x} |u|^2  \;. 
\end{align*}
Noting that $
| a^\dag u|^2 = | a u|^2 -2 \chi W \dpt{|u|^2}{x} $, and using assumption $ii)$ together with $\zg^+ < \zg^-$, we obtain $A \le - \langle u, Y u \rangle$. Combining this estimate with standard domain considerations for the respective operators, the results of~\cite{Fagnola_Rebolledo_01} yield the desired conclusion.
\endproof

  \section{Wigner-Lindblad equation }\label{sec_WL_eq}
  The Wigner transform maps the density matrix kernel onto phase-space functions. The Wigner quasi-probability distribution function associated with the density matrix $\zr$ is defined as
  \begin{align}
  	f \left(x,p\right)\doteq &\frac{1}{(2\pi)^{n}}	\int_{\mathbb{R}^{n}} \zr \left(x+\frac{\hbar{ \zh}}{2} ,x-\frac{\hbar{ \zh}}{2}\right)  e^{-ip \zh}\dif { \zh}\;.\label{def_W_f}
  \end{align}
  The evolution equation for the Wigner function can be systematically derived through the formalism of Weyl quantization. The Wigner transform provides an algebra isomorphism between the space of trace-class operators and the space of phase-space symbols equipped with the Moyal product~\cite{Folland_89}. We define the inverse Weyl transform as in Eq.~\eqref{def_W_f}, $f = \text{Op}^{-1} (\zr)$, which maps a trace-class operator to its corresponding phase-space symbol. With this definition, the operator product $AB$ is mapped to the Moyal product $A\#B = \text{Op}^{-1}(AB)$, where the Moyal product is defined by the integral 
  \begin{align}
  	A \# B
  	=&    \frac{ 1}{(2\pi)^{2n}}   \int_{\mathbb{R}^{2n}}  A \left(x -\frac{\hbar \zh}{2},p+ \frac{\hbar\zm}{2} \right) B \left(x',p' \right) e^{i\mu(x-  x')+i\zh(p -p') } \textrm{d}\mu \dif \zh   \dif x' \textrm{d} p'\;. \label{moyal_prod}
  \end{align}
  The evolution equation for the Wigner function can be compactly formulated as a von Neumann-type equation within the Moyal algebra. By applying the $\text{Op}^{-1}$ map to the Lindblad master equation, we obtain the full quantum Wigner--Lindblad problem:
  \begin{align} 
  	&  \dpp{f  }{t} + \frac{p}{m}\cdot   \nabla_x f  + \Theta_{U}[   f ]  -T(f)=0 \label{Wig_Lind}    
  \end{align} 
  The pseudodifferential operator $\Theta_U[f]$ is defined as
  \begin{align}
  	\Theta_U [f](x,p)  \doteq&    \frac{ 1}{(2\pi)^{n}}   \int_{\mathbb{R}^{2n}} \frac{1}{i\hbar}\left[U \left(x +\frac{\zh\hbar}{2}  \right)-U \left(x -\frac{\zh\hbar}{2}  \right)\right] f \left(x,p' \right) e^{- i\zh(p -p') }    \dif \zh   \dif p'\;, \label{Theta_pm}
  \end{align}
  and the collision operator for the Wigner function is defined as
  \begin{align} 
  	&T(f)\doteq 	\zg^+ \left( \overline{a}  \# f \# a -\frac{1}{2}\{ a  \# \overline{a}  , f  \}_{\#}\right)  + \zg^- \left(  a \# f \# \overline{a}  -\frac{1}{2}\{ \overline{a} \# a , f  \}_{\#}\right) \; . \label{WL_all_ord}
  \end{align}
Recently, a similar approach based on creation and annihilation operators associated with the harmonic potential was presented in~\cite{Steuernagel_25}. 
The Moyal product admits a systematic semiclassical expansion in powers of Planck's constant $\hbar$, given by $A \# B = A B + \frac{i\hbar}{2} \{ A, B \} - \frac{\hbar^2}{8} \Lambda^2(A, B) + O(\hbar^3)$, where $\{ A, B \} = \nabla_x A \cdot \nabla_p B - \nabla_p A \cdot \nabla_x B$ denotes the classical Poisson bracket, and $\Lambda^2$ represents a second-order differential operator. This asymptotic expansion establishes a precise bridge between quantum phase-space evolution and classical Liouville dynamics, yielding the leading-order quantum corrections in the semiclassical regime.

Due to the complexity of the full quantum dynamical evolution, in the spirit of the Fokker--Planck approach, we simplify the full quantum Wigner--Lindblad equation by truncating the $T(f)$ operator to second-order differential terms. This truncation aligns with the semiclassical expansion of the Moyal commutator, which is the standard procedure in the Wigner formulation for obtaining leading quantum corrections to classical dynamics. From a physical perspective, thermalization toward equilibrium is essentially a classical phenomenon in which quantum effects manifest as subleading corrections. Conversely, we retain a full quantum description for the external potential, which governs the coherent quantum evolution of the system. In doing so, we implicitly assume that our model applies to regimes where coherent quantum dynamics dominate, while environmental interactions and energy exchange with the thermal bath act as second-order effects whose primary role is to drive the system toward local thermal equilibrium.
By expressing $a = g(x,p) + i h(x,p)$ in terms of its real and imaginary parts, the Lindblad collision term to second order in the semiclassical expansion becomes 
\begin{align} 
	&T(f)= \text{Div}_{x,p}   ( F )= \nabla_{x}   \cdot   F_x + \nabla_{p} \cdot   F_p \;,   \label{WL_sec_ord}
\end{align}
where
\begin{align*} 
	F_x= &\sum_{i=1}^n \left\{   D \left[   \left( \left|\nabla_p g_i   \right|^2 +   \left|\nabla_p h_i   \right|^2  \right) \nabla_x f   +\Gamma(g,h)   \nabla_p f  \right] + \zg   \left[ -(\nabla_p g_i)   h_i +( \nabla_p h_i ) g_i \right] f   \right\}  \\
	F_p= & \sum_{i=1}^n \left\{   D \left[   \left( \left|\nabla_x g_i   \right|^2 +   \left|\nabla_x h_i   \right|^2  \right) \nabla_p f   +\Gamma(g,h) \nabla_x f  \right]+\zg   \left[ (\nabla_x g_i)   h_i -( \nabla_x h_i ) g_i\right] f   \right\}   \;,
\end{align*}
and we have defined $ \Gamma (g,h) \doteq     - \left(\nabla_p g_i \cdot \nabla_x g_i   +\nabla_p h_i \cdot \nabla_x h_i \right)$, with $D=\frac{\hbar}{2}( \zg^++\zg^-)$, $\zg= \zg^-- \zg^+$. 
Using the annihilation symbol derived in the previous section, $a = W+i\frac{p}{\sqrt{2m}}$, the collision operator simplifies to
%
\begin{align*} 
	T(f)= &   \frac{nD}{2m} \Delta_x f   +\frac{\zg}{\sqrt{2m}}   \text{div}_x \left(   W   f \right)    +D \left(\sum_{i=1}^n   \left|\nabla_x W_i   \right|^2   \right)   \Delta_p f    +\frac{\zg}{\sqrt{2m}}    \sum_{i=1}^n    \left(  \nabla_x W_i   \right) \cdot \nabla_p \left( p_i f   \right)\;. 
\end{align*}
	For harmonic confinement, $W=\frac{\zo \circ x}{\sqrt{2}}  $, the collision operator reduces to the Fokker--Planck operator, as expected:
	\begin{align*} 
		T(f)= &   \frac{n D}{2m} \Delta_x f   +\frac{D}{2}\left(\sum_i   \zo_i^2    \right) \Delta_p f      +\frac{\zg }{2\sqrt{m}}\bigg[ \text{div}_x \left[ \left( \zo \circ x   \right) f  \right] + \text{div}_p \left[ \left( \zo \circ p   \right) f  \right]   \bigg] \;. 
	\end{align*}
	The coefficient $\gamma$ represents the rate of energy exchange and can be expressed as $	\gamma =\zg^--\zg^+ = 2 \Gamma    \frac{\hbar} {\Delta E }$, where $\Gamma$ denotes the collision frequency of the particles with the external bath and $\Delta  E$ represents the mean energy exchanged per collision.
	A comparison with the Fokker--Planck collision operator in the one-dimensional case, which we denote by $T_{\text{FP}}(f) $, provides a straightforward procedure to fix the remaining coefficients. We have
	\begin{align*} 
		T_{FP}(f) =& D_x   \dsp{f}{x} +  D_p   \dsp{f}{p} + 2\zg_{FP}  \dpp{(pf)}{p} \\   
		T(f)= &   \frac{D}{2m} \dsp{f}{x}  +\frac{\zg}{\sqrt{2m}}  \dpp{ \left(   W   f \right)  }{x}  +D  \left(\dpt{W}{x}   \right)^2   \dsp{f}{p} +\frac{\zg}{\sqrt{2m}}  \dpt{W}{x}     \dpp{\left( p f   \right)}{p}\;. 
	\end{align*}
	For the FP operator, we have $ 2 \gamma_{FP}=\Gamma$. Assuming local equilibrium described by the Maxwell--Boltzmann distribution (detailed balance) implies that $ \frac{D_p}{\zg_{FP}}=  2 m k_B T_B$, where $T_B$ denotes the temperature of the bath. An estimate of the spatial diffusion coefficient can be obtained using the Lindblad relation $D_pD_x = \zeta \frac{\hbar^2\zg^2_{FP}}{4}$ \cite{Sparber_02,Arnold_08}, where we have introduced the parameter $\zeta\geq 1$ to ensure that the Lindblad inequality is always satisfied.
	Our method simplifies to the standard Fokker--Planck model when $W = x \sqrt{\frac{m}{2}} \frac{\Delta E }{\hbar} $. Comparing the respective coefficients, we obtain $\gamma  = 4m  \gamma_{FP} \sqrt{\frac{D_x}{D_p}} $, $D = D_x  2 m$, and $\frac{\Delta E}{\hbar} = \sqrt{\frac{D_p}{D_x}}\frac{1}{m}  $. Note that for harmonic confinement, $W= \frac{\zo x}{\sqrt{2}} $, with $U_{ext}= \frac{\zo^2}{2}x^2$, we obtain $\zo= \sqrt{m} \frac{\Delta E }{\hbar} =\sqrt{\frac{D_p}{mD_x}} $.

 \section{Numerical results}\label{sec_num_res}
 
Numerical studies of Wigner dynamics are mostly restricted to 1D systems (2D in phase space), with only a few examples developed for higher-dimensional spaces, which are typically limited to specific cases~\cite{Xiong_23}. 
We test our method on textbook cases by comparing the long-time behavior of the solution to the Wigner-Lindblad problem with the FP approximation. Specifically, we analyze three scenarios. The first two, formulated in 2D and 4D phase space, investigate textbook cases of anharmonic confinement that serve as prototypical models in molecular dynamics. As a final example, we consider the charge flux through a resonant tunneling device.

 \subsubsection*{1D Anharmonic Trap}
\begin{figure}[!h]
 	\begin{center}
 		\includegraphics[width=0.8\textwidth]{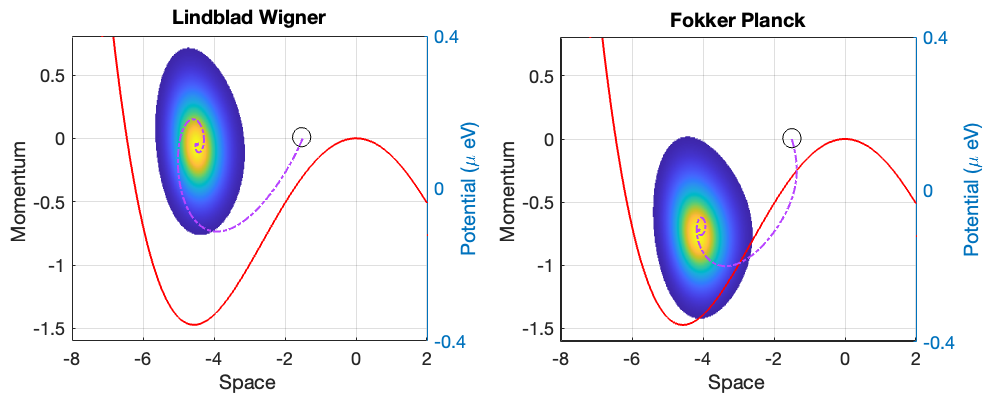} 
 	\end{center}
 	\caption{
 		Phase-space dynamics of a trapped atom. Left panel: Wigner--Lindblad model. Right panel: Quantum Fokker--Planck model. The density distribution displays the quantum state at the final time $T = \qty{1.0}{\milli\second}$. The circular marker denotes the initial phase-space mean coordinate $(\langle x(0) \rangle, \langle p(0) \rangle)$. Solid red curve: external potential profile $U_{\text{ext}}$, plotted on the same vertical axis as the momentum. Dashed violet curve: time evolution $(\langle x(t) \rangle, \langle p(t) \rangle)$. }
 	\label{fig_comp_WL_FP} 
 \end{figure}

 As a first example, we investigate the dynamics of a single localized particle, specifically a strontium atom ($^{87}\text{Sr}$) in our simulations, trapped within a quartic double-well potential given by 
 $ 	U(x) = u_4 x^4 - u_2 x^2$, where $u_4 > 0$ and $u_2 > 0$ represent the quartic confinement and quadratic parameters, respectively. This potential forms two symmetric traps with a well depth of approximately $\qty{0.5}{\micro\eV}$ and a spatial separation of around $\qty{10}{\nano\meter}$. 
 The initial state of the particle is prepared as a localized Gaussian wave packet. In the quantum phase-space formulation, this corresponds to the initial Wigner distribution $
 f_0 = \frac{ \sqrt{2mk_B T_B}}{\Delta_x\pi}\exp\left[ -\frac{(x - x_0)^2}{\Delta_x^2}  - \frac{1}{k_BT_B}\frac{p^2}{2m}  \right]$. 
 %
 The particle is coupled to a thermal bath characterized by a temperature $T_{\text{B}}$. Figure~\ref{fig_comp_WL_FP} compares the time evolution of the atomic quantum phase-space distribution obtained using the Wigner--Lindblad model and the standard Fokker--Planck relaxation mechanism. 
 
 Given its initial coordinates $(\langle x(0) \rangle, \langle p(0) \rangle)= (-1.5,0)$, the atom is located inside the left well, near the central potential barrier. This corresponds to a physical regime in which the particle is bound within the left trap with a total energy close to the barrier height, lacking sufficient kinetic energy to escape the well classically.
 The solid red curve in Fig.~\ref{fig_comp_WL_FP} represents the potential energy profile $U$ as a function of position. To facilitate visual comparison between the atomic coordinates and the potential landscape, both the particle momentum and the external potential energy are plotted along the same vertical axis.
 In the absence of energy dissipation, the atom undergoes undamped, coherent back-and-forth oscillations within the well. In the figure, a circular marker indicates the initial phase-space centroid of the wave packet, $(\langle x(0) \rangle, \langle p(0) \rangle)$. Coupling to a thermal bath at a temperature lower than the well depth induces energy dissipation, driving atomic relaxation and localization toward the potential minimum.
 To trace the dynamical evolution of the atomic density, the dashed violet curve illustrates the time evolution of the mean position and momentum, $(\langle x(t) \rangle, \langle p(t) \rangle)$. The density profile reflects the state of the system at the final time $T = \qty{1.0}{\milli\second}$.
 The left panel of Fig.~\ref{fig_comp_WL_FP} displays the solution of the Wigner--Lindblad model, whereas the right panel shows the solution of the quantum Fokker--Planck framework. A clear qualitative discrepancy emerges in the long-time asymptotic behavior of the two solutions. In the WL case, the wave packet is driven toward the potential minimum of the left well, and the particle's mean momentum vanishes at long times, as physically expected. In contrast, under the FP framework, the Gaussian packet oscillates around a spatial position offset from the well minimum, and its long-time mean momentum remains markedly non-zero. These results indicate that, in the present scenario, the FP model fails to provide a physically accurate description of the relaxation mechanism, thereby motivating the use of alternative models to properly capture the atomic thermalization process.
 In our simulations, we use the following parameters: quartic coefficient $u_4=\qty{1.03e-03}{\micro\electronvolt \nano\meter^{-4}}$, quadratic coefficient $u_2=\qty{4.31e-02}{\micro\electronvolt \nano\meter^{-2}}$, initial mean position $x_0=\qty{-1.5}{\nano\meter}$, initial mean momentum $p_0=0$, initial spatial spread $\Delta_x=\qty{0.63}{\nano\meter}$, bath temperature $T_{\text{B}}= \qty{100}{\micro\kelvin} $, friction $\zg_{FP}=\qty{0.8}{\micro\second^{-1}}$, spatial and momentum diffusion coefficients, $D_x=\qty{0.04}{\nano\meter^{2}\micro\second^{-1}} $ and $D_p/\hbar^2=\qty{7.7e4}{\nano\meter^{-2}\micro\second^{-1}} $, respectively.

 \subsubsection*{2D Anharmonic Trap}
 
 As a second application of the Wigner--Lindblad model, we consider the dynamics of an electron in a two-dimensional (2D) potential landscape, corresponding to a four-dimensional (4D) phase-space simulation. In this test case, we employ an external potential similar to the 1D scenario,
 $
 U(x, y) = u_4 x^4 - u_2 x^2 + v_2 y^2$,  
 which combines a fourth-order double-well profile along the $x$-direction with harmonic confinement along the $y$-direction.
 At $t = 0$, the particle is prepared as a Gaussian wave packet centered at phase-space coordinates $(r_0, p_0)$, with spatial and momentum spreads $\Delta_r$ and $\Delta_p$, respectively.  
 \begin{figure}[!h]
 	\begin{center}
 		\includegraphics[width=0.8\textwidth]{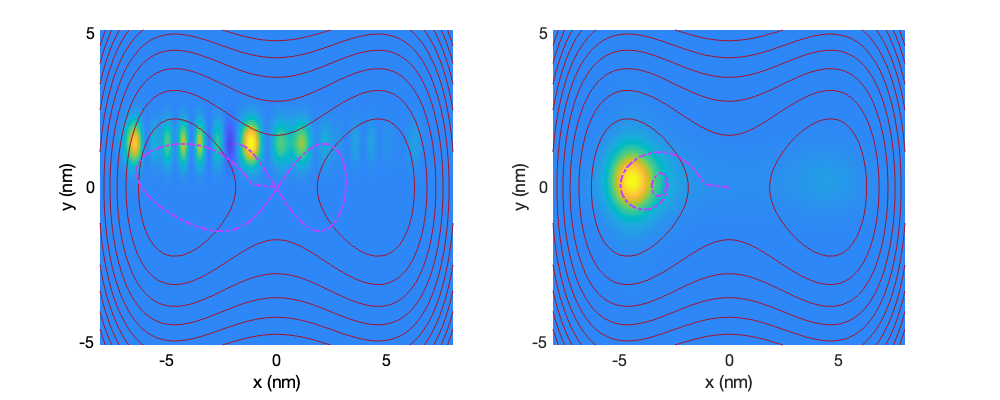} 	
 		\includegraphics[width=0.8\textwidth]{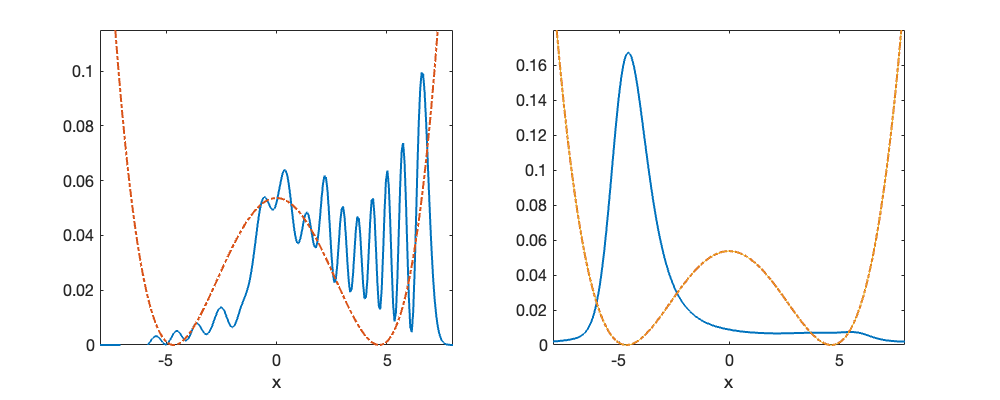}
 	\end{center}
 	\caption{
 		Comparison between undamped, fully coherent dynamics (left panels) and dissipative dynamics under the Wigner--Lindblad model (right panels). Upper panels: Snapshots of the state evolution at $t = 400$, where the black circle indicates the initial phase-space centroid, the dashed violet curve traces the center-of-mass trajectory, and red lines show potential contour lines. Lower panels: Integrated particle density profiles along the $x$-direction at an intermediate time $t = 230$.} \label{fig_2D_01}
 \end{figure}
 
 The simulation results are illustrated in Fig.~\ref{fig_2D_01}, where we compare the undamped, fully coherent dynamics (left panels) with the dissipative dynamics governed by the Wigner--Lindblad relaxation mechanism (right panels). The initial phase-space position of the particle is indicated by a black circle, and the initial momentum is directed toward the upper-left quadrant. 
 
 The upper panels show snapshots of the state evolution at $t = 400$. The dashed violet curve represents the center-of-mass trajectory, while the red curves display the contour lines of the external potential. In the undamped case, the wave packet explores both the left and right potential minima and undergoes significant broadening, forming a coherent quantum state delocalized across both wells. In contrast, the inclusion of damping achieves efficient spatial localization, keeping the particle well confined within a single potential well.
 The lower panels display the particle density along the $x$-direction at an intermediate time $t = 230$. This snapshot captures a state where the undamped wave packet bounces off the right barrier, overcomes the central barrier, and propagates back toward the left well. The density profile exhibits a clear interference pattern characteristic of quantum reflection in coherent states. Conversely, in the right panel (with damping), the particle density remains strongly localized within the left well, with only a minor probability density residing in the right well. All numerical solutions were obtained using the 4D Wigner code~\cite{WIG2D}, employing a dedicated spectral split-operator routine to integrate the relaxation terms into the time-evolution algorithm.
 In our simulations, we use the following parameters: quartic coefficient $u_4=\qty{1.1}{\milli\electronvolt \nano\meter^{-4}}$, quadratic coefficient $u_2=\qty{49.5}{\milli\electronvolt \nano\meter^{-2}}$, initial mean position $r_0=(x_0,y_0)=(-1,0)\;\qty{}{\nano\meter}$, initial mean momentum $p_0=(p_{x_0},p_{y_0})=(-2,2)\;\qty{}{\nano\meter^{-1}}\hbar $, initial spatial spread $\Delta_x=\qty{2.5}{\nano\meter}$, $\Delta_p=\qty{1}{\nano\meter^{-1}}\hbar $, friction $\zg_{FP}=\qty{11.5}{\nano\second^{-1}}$, and spatial and momentum diffusion coefficients $D_x=\qty{11.3}{\nano\meter^{2}\nano\second^{-1}} $ and $D_p/\hbar^2=\qty{2.2e7}{\nano\meter^{-2}\nano\second^{-1}} $, respectively.


\subsubsection*{Resonant tunneling diode}

\begin{figure}[!h]
	\begin{center}
		\includegraphics[width=0.45\textwidth]{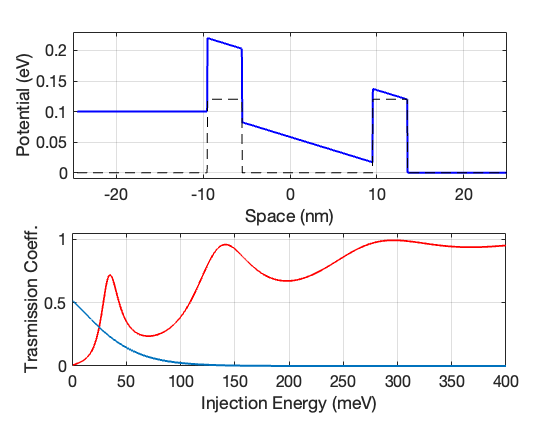} 
		\includegraphics[width=0.45\textwidth]{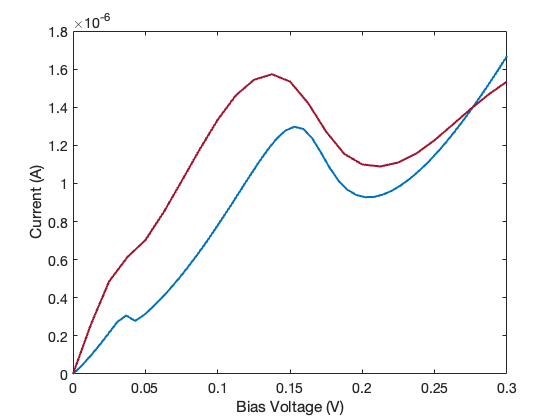} 	
		\end{center}
	\caption{Top-left panel: Potential profile of the InAs/InP double-barrier RTD unbiased (black dashed curve) and under an applied bias of $\qty{0.1}{\volt}$ (blue solid curve). Lower subpanel: Red curve shows the plane-wave electron transmission probability across the unbiased device as a function of injection energy; blue curve denotes the equilibrium Fermi--Dirac distribution. Right panel: Current--voltage diagram obtained with the Wigner code (red curve) and the collisionless Green's function model (blue curve).}
	\label{fig_RTD} 
\end{figure}

As a final application of our particle--bath interaction model, we simulate the trapping time and the electron flux through a solid-state resonant tunneling device. We consider a double-barrier heterostructure consisting of an undoped InAs substrate with two $\qty{4}{nm}$-thick InP barrier layers separated by a $\qty{15}{nm}$ InAs well layer. The potential profile of the resonant tunneling diode (RTD) structure is depicted in Fig.~\ref{fig_RTD} for the unbiased case (black dashed curve) and under an applied bias of $\qty{0.1}{\volt}$ (blue solid curve). The height of the potential barriers is set to $\qty{0.12}{\electronvolt}$. The double-barrier structure forms a quantum well containing localized states, referred to as resonant levels. When the energy of the electrons matches a resonant level, electrons tunnel efficiently through the barriers. To clarify the electronic structure of the device, the top-left panel shows the plane-wave electron transmission probability across the unbiased device as a function of the particle injection energy. Each peak corresponds to a quasi-resonant state bound between the two thin barriers. To allow direct comparison with the thermal energy of the contact electrons, the lower subpanel displays the equilibrium distribution function of the injection scattering state, $f_{\text{eq}} =f_{FD}(E;T,\mu) $, where $f_{FD}(E;T,\mu) $ denotes the Fermi--Dirac distribution $f_{FD}(E;T,\mu)  = \left(1 + e^{\frac{E-\mu }{k_\text{B} T}}\right)^{-1}$.
As the potential difference between the contacts is varied, the current flowing through the device exhibits characteristic non-monotonic behavior, displaying negative differential resistance (NDR) in its current--voltage ($I$--$V$) characteristics. The physical mechanism underlying the current--voltage behavior of the device is well understood and has been widely discussed in the literature (see, e.g., \cite{Kouwenhoven_01,Wagner_19,Esseni_19,Sakurai_13,Sakurai_14,Hendrickx_20,Bjork_02}). In the right panel of Fig.~\ref{fig_RTD}, we compare the $I$--$V$ diagram obtained with the Wigner code (red curve) and the collisionless Green's function model (blue curve) \cite{Jauho_94,Gurvitz_96}. The two models show good agreement. In our simulations, we use the following parameters: collision rate $\Gamma = \qty{1}{\nano\second^{-1}} $, $\zeta=1$, bath temperature $T_{B}=\qty{300}{\kelvin}$, chemical potential $\mu=\qty{5 e-3}{\electronvolt^{-1}}$, and Fokker--Planck momentum and spatial diffusion coefficients $D_p = \Gamma  m T_B$ and $D_FP_x = \hbar^2/ \Gamma^2/D_p$, respectively (see discussion in Sec.~\ref{sec_WL_eq}).

\begin{figure}[!h]
	\begin{center}
 		\includegraphics[width=0.45\textwidth]{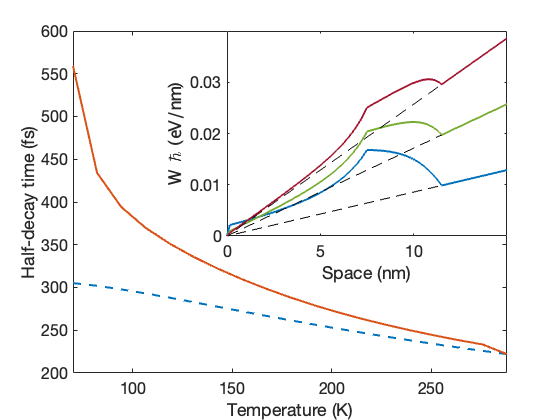}   
	\end{center}
	\caption{Half-decay time of the trapped electron density as a function of the bath temperature, comparing the Wigner--Lindblad model (red curve) and the Fokker--Planck model (blue curve). Subpanel: Effective potential profile $W(x)$ for $x \ge 0$ at temperatures of $100~\text{K}$ (dashed blue curve), $200~\text{K}$ (green curve), and $300~\text{K}$ (red curve). The dashed straight lines depict the corresponding linear approximation of the Fokker--Planck model.}
	\label{fig_trap_time} 
\end{figure}
%
We investigate the trapping time of the electron density inside the well and compare the results obtained via the Fokker--Planck approach with those from the Wigner--Lindblad formulation in the high-collision regime. Figure~\ref{fig_trap_time} illustrates how the trapping time is estimated by the two models. We consider the evolution of a thermalized charge with a Gaussian spatial profile represented by the Wigner function $f(x,p)=e^{-\frac{x^2}{\Delta_x} }f_{FD}\left(\frac{p^2}{2m}; T_{B},\mu\right)$, with $\Delta_x= \qty{10}{\nano\meter}$, initially trapped inside the central potential well in a strongly collisional regime characterized by a collision rate $\Gamma= \qty{e2}{\nano\second^{-1}}$.
We estimate the trapping time by the half-decay time of the trapped density for different temperatures of the bath. In the figure, the red curve refers to the Wigner--Lindblad model and the dashed blue curve to the Fokker--Planck model. The two models agree at high temperatures, whereas at lower temperatures the Wigner--Lindblad model predicts a longer decay time. 
The reason for this discrepancy can be understood by analyzing the shape of the potential $W$ at different temperatures. In the subpanel of Fig.~\ref{fig_trap_time}, we show the potential $W$ for $x\geq 0$ ($W(-x)=-W(x)$) corresponding to temperatures of $100~\text{K}$ (blue curve), $200~\text{K}$ (green curve), and $300~\text{K}$ (red curve). The dashed straight lines correspond to the Fokker--Planck case (i.e., replacing $W$ with the corresponding straight line causes the Wigner--Lindblad problem to coincide with the Fokker--Planck problem). We observe that as the temperature increases, the potential $W$ approaches the associated straight line; consequently, the half-decay time estimates of the Wigner--Lindblad and Fokker--Planck approaches coincide.

 \section{Conclusion}
 
In this work, we have developed a kinetic framework for open quantum systems that incorporates the Lindblad theory of particle--bath interaction into the Wigner phase-space formulation. We constructed a set of potential-dependent creation and annihilation operators. A second-order semiclassical expansion of the Moyal star product led to a generalized Wigner--Lindblad kinetic model featuring position-dependent friction and diffusion mechanisms alongside full coherent quantum potential terms.
From a mathematical perspective, we proved the well-posedness of the Lindblad dynamics, establishing the existence of a trace-preserving $w^*$-continuous semigroup and a unique steady state. Our framework provides a promising theoretical tool for investigating dissipative quantum transport and relaxation phenomena in nanodevices and molecular systems.
Numerical simulations validated the physical accuracy and versatility of our approach. In quartic double-well potentials, the Wigner--Lindblad model effectively restores physical thermalization toward potential minima, resolving the persistent non-zero momentum artifacts observed in traditional quantum Fokker--Planck models. In multi-dimensional landscapes (4D phase space), the formulation accurately captures dissipative quantum state localization and decoherence during wave-packet propagation. Finally, the model successfully describes resonant transport, current--voltage characteristics, and electron trapping kinetics in InAs/InP double-barrier diodes.

 \section*{Acknowledgments}
 	The work has been developed under the auspices of GNFM (INdAM). 
 	
 \appendix
 
 \section{Proof of Theorem~\ref{teor_well_pos-LP}}\label{App_proof_Th_wel_pos}
 We define the auxiliary operator $G= -i  H +G_0$, where 
 \begin{align}
 	G_0=& - \frac{1}{2} \left(\zg^- a^\dagger a + \zg^+ a a^\dagger \right) =   \frac{D}{\hbar}  (I E_0-H)- \zg^+ \chi \dpt{W}{x}  
 	\label{def_G}
 \end{align}
 with domain $D(G) = H^2\left(\mathbb{R}\right)$, where we have defined $\chi = \hbar\frac{1}{\sqrt{2m}}$ and $D = \hbar \frac{\zg^-+\zg^+}{2}$. By treating the term $\dpt{W}{x}$ as a bounded perturbation, it is straightforward to show that the operator $G_0$ is self-adjoint. We start with the following lemma.
 
 \begin{lemma}[Contractivity]\label{lemma_cont_G}
 	The operator $G$ is the infinitesimal generator of a contraction semigroup $(P_t)_{t \ge 0} = (e^{tG})_{t \ge 0}$ on $L^2\left(\mathbb{R}\right)$.
 \end{lemma} 
 \proof 
 Since $G_0$ is self-adjoint, by the Lumer--Phillips theorem, it is sufficient to prove dissipativity to ensure that $G$ generates a $C_0$-contraction semigroup.
 We have $ \Re \langle x , G x\rangle  \leq -  \frac{D}{\hbar} \left(  \langle H x,x \rangle -   E_0    \right)\leq 0$, where we used assumption $ii)$ of Theorem~\ref{teor_well_pos-LP}.  
 \endproof
 
 We denote by $\mathcal{T}_t \colon \mathcal{J}_1(L^2\left(\mathbb{R}\right)) \to \mathcal{J}_1(L^2\left(\mathbb{R}\right))$ the evolution semigroup associated with the LP. The time evolution of the expectation value of an observable $A \in B( L^2\left(\mathbb{R}\right))$ for a state initially represented by the density matrix $\zr$ is given by $\operatorname{Tr}\!\left( \mathcal{T}_t (\rho) A \right) = \operatorname{Tr}\!\left( \rho \, \mathcal{T}_t^* (A) \right)$, where the right-hand side refers to the associated Heisenberg picture, with semigroup $\mathcal{T}_t^* \colon B( L^2\left(\mathbb{R}\right)) \to B( L^2\left(\mathbb{R}\right))$. In the present context, motivated by the duality ${\mathcal{J}_1(L^2\left(\mathbb{R}\right))}^* = B( L^2\left(\mathbb{R}\right))$, the semigroup $\mathcal{T}_t$ is referred to as the predual semigroup. It is more convenient to conduct the existence analysis of the evolution semigroup in the space of bounded operators. In particular, one can exploit the fact that the $w^*$-continuity of $\mathcal{T}_t^*$ on $B( L^2\left(\mathbb{R}\right))$ implies the strong continuity of $\mathcal{T}_t$ on $\mathcal{J}_1(\mathcal{H})$ \cite{Fagnola_Rebolledo_03}. If $\mathcal{L}$ and $\mathcal{L}^*$ denote the generators of $\mathcal{T}_t$ and $\mathcal{T}_t^*$, respectively, the duality between the Heisenberg and Schrödinger pictures transfers to the generators via $\operatorname{Tr}\!\left( \mathcal{L}  (\rho) A \right) = \operatorname{Tr}\!\left( \rho \, \mathcal{L}^* (A) \right)$.
 
 The generator $\mathcal{L}^*$ of the semigroup $\mathcal{T}_t^*$ is associated with the sesquilinear form $\mathcal{L}^*(A)(u,v)$, with domain $D(G) \times D(G)$ and $A\in B(L^2\left(\mathbb{R}\right))$, given by
 \begin{align*}
 	\mathcal{L}^* (A)(u,v) =&  i \langle v, \left[H, A  \right] u\rangle   -\frac{\zg^-}{2} \langle  v, \{ a^\dag a , A\}  u  \rangle   -\frac{\zg^+}{2} \langle  v, \{ a a^\dag , A\}  u  \rangle + \zg^- \langle a u , A a v \rangle+ \zg^+ \langle a^\dag u , A a^\dag v \rangle
 	\\
 	=& \langle G v, A u  \rangle +\langle  v, A G u  \rangle + \langle L_- u , A L_- v \rangle + \langle L_+ u , A L_+ v \rangle\;, 
 \end{align*}
 with $L_+ = \sqrt{\zg^+}\; a^\dag$ and $L_- = \sqrt{\zg^-} \; a$. We have used the Heisenberg--Schrödinger correspondences $\mathcal{L}(\rho) \doteq a \rho a^\dag \rightarrow \mathcal{L}^*(A) = a^\dag A a$, $\mathcal{L}(\rho) = \{ a^\dag a ,\zr \} \rightarrow \mathcal{L}^*(A) = \{ a^\dag a ,A \}$ and $\mathcal{L}(\rho) = [H ,\zr] \rightarrow \mathcal{L}^*(A) =- [H , A]$. We highlight the relation between $G$ and the ladder operators, $G^\dag + G =2 G_0= - \sum_{i=\pm} L_i^\dag L_i$.
 
 
 We prove the existence of the Markovian semigroup stated in Theorem~\ref{teor_well_pos-LP}. We recall the following result (see Corollary 4.5 in \cite{Chebotarev_Fagnola_98}):
 
 \begin{theorem}[Chebotarev--Fagnola]\label{th_Cheb_Fagn_01} Assume that $1)$ the operator $G$ is the infinitesimal generator of a strongly continuous contraction semigroup $(P(t))_{t\geq 0}$ in $L^2(\mathbb{R})$. The domain of the operator $a$ contains the domain $D(G)$ of the operator $G$. For all $u, v \in  D(G)$, we have $\langle v, Gu\rangle +\langle Gv, u\rangle +\sum_i \langle L_i v, L_i u\rangle =0.$
 $2)$ There exist positive self-adjoint operators $C$ and $\Phi$ in $L^2(\mathbb{R})$ and a core $\mathcal{D}$ for $G$ such that 
 $D(G)= D(C)$ and for all $u\in D(G)$ there exists a sequence $(u_n)_{n\in \mathbb{N}} \in \mathcal{D}$ such that both $(G u_n)_{n\in \mathbb{N}}$ and $(C u_n)_{n\in \mathbb{N}}$ converge strongly. $3)$ There exists a positive constant $b$ such that
 $2 \Re \langle G u, C u\rangle 
 + \langle   a u, C   a u\rangle 
 \leq b \langle u, Cu\rangle$, for every $ u \in \mathcal{D}$.
 $4)$ $a ( \mathcal{D}) \subseteq D(C)$.
 $5)$ $\mathcal{D} \subseteq D(\Phi)$, and for every $u\in \mathcal{D}$ we have $- 2 \Re \langle u, Gu\rangle 
 =\langle  u, \Phi u\rangle \leq   \langle  u, C u\rangle$. 
 Then the minimal quantum dynamical semigroup is conservative.
 \end{theorem}
 
 
 We apply Theorem~\ref{th_Cheb_Fagn_01} by setting $ C \doteq  - 2 G_0 +I\zl=  -2G -i H +I\zl$, with $\zl>0$, and $\Phi =-2G_0$ with $D(\Phi)=D(C) = H^2(\mathbb{R})$. The core for $G$ is taken as $\mathcal{D}=C_0^\infty(\mathbb{R})$. Condition 1 is verified by Lemma~\ref{lemma_cont_G}, while Condition 2 is trivially satisfied. 
 Using the commutator relation~\eqref{comm_adag_a} together with assumption $ii)$ of Theorem~\ref{teor_well_pos-LP}, straightforward calculations yield
 \begin{align*}
 A\doteq  & 	2\text{Re}	\langle G u, C u \rangle +\sum_i	\langle L_i u, C L_i u \rangle \\
 =&  2\chi\left( \zg^-  +  \zg^+ \right) \int  \left(-\zg^- | a u|^2   +\zg^+   | a^\dag u|^2 \right)\dpt{W}{x} \dif x - 4\chi^2   \zg^+\zg^-    \int W   \dst{W}{x} |u|^2  .
 \end{align*}
 Using the identity $| a^\dag u|^2 = | a u|^2 -2 \chi W \dpt{|u|^2}{x}$, we obtain
 \begin{align*}
 A= &   2\chi  \left( -(\zg^-)^2  + ( \zg^+)^2 \right)\int | a u|^2 \dpt{W}{x}  \dif x+ 4 \chi^2   \left( \zg^+  \right)^2 \int |u|^2 \dpt{}{x} \left(W    \dpt{W}{x} \right)\dif x \\& + 4\chi^2   \zg^+\zg^-    \int \left| \dst{W}{x}\right|^2 |u|^2 \dif x\leq  K  \| u \|^2_{L^2}   \;,
 \end{align*}
 where $K= 4 \chi^2 (\zg^-)^2\sup_{x\in\mathbb{R}} \left(W^2+ \left| \dpt{W}{x}\right|^2+\left| \dst{W}{x}\right|^2 \right)$. Here we used assumption $ii)$ of Theorem~\ref{teor_well_pos-LP} and the fact that $\zg^+ < \zg^-$. 
 Since 
 \begin{align*}
 & 	\langle  u, C u \rangle =  \zl   \int |   u|^2 \dif x+   \zg^-   \int | a u|^2 \dif x + \zg^+\int | a^\dag u|^2\dif x,
 \end{align*}
 we have $2\text{Re}\langle G u, C u \rangle +\sum_i\langle L_i u, C L_i u \rangle \leq b \langle u, C u \rangle$, so Condition 3 is satisfied. Condition 5 follows from similar considerations since $G=-i H -\frac{1}{2}\Phi$, $C=\Phi+I$, and $-2 \text{Re}\langle G u, u \rangle = \langle u, \Phi u \rangle = \langle u, C u \rangle -\zl \|u\|^2$. Regarding Condition 4, we note that since $W \in H^2_{\text{loc}}$, we have $a(C^\infty_0(\mathbb{R}))\subseteq H^2 (\mathbb{R}) = D(C)$.
 
 We can thus conclude the existence of the minimal Markovian quantum dynamical semigroup. The remaining parts follow from Proposition III.1 of \cite{Fagnola_Rebolledo_01} (i.e., under the assumption of Condition 1, the minimal QDS is Markovian, $\mathcal{T}_t^{\text{min}} (I)=I$, if and only if $\mathcal{T}_t^{\text{min}} (I)=I$ is the unique $w^*$-continuous family of positive contractive maps on $B(L^2(\mathbb{R}))$ satisfying the LP for all positive $x \in B(L^2(\mathbb{R}))$ and all $t>0$).
  \endproof






\end{document}